# Quantitative measurements of transverse thermoelectric generation and cooling performances in $SmCo_5/Bi_{0.2}Sb_{1.8}Te_3$-based artificially tilted multilayer module

Masayuki Murata[a], Fuyuki Ando[b], Takamasa Hirai[b], Hiroto Adachi[c], and Ken-ichi Uchida[b,d]

[a] Research Institute for Energy Conservation, National Institute of Advanced Industrial Science and Technology, Tsukuba, Japan;

[b] Research Center for Magnetic and Spintronic Materials, National Institute for Materials Science, Tsukuba, Japan;

[c] Research Institute for Interdisciplinary Science, Okayama University, Okayama, Japan;

[d] Department of Advanced Materials Science, Graduate School of Frontier Sciences, The University of Tokyo, Kashiwa, Japan

**CONTACT**

Masayuki Murata, m.murata@aist.go.jp, Research Institute for Energy Conservation, National Institute of Advanced Industrial Science and Technology, 1-1-1 Umezono, Tsukuba, Ibaraki, Japan

Ken-ichi Uchida, UCHIDA.Kenichi@nims.go.jp, Research Center for Magnetic and Spintronic Materials, National Institute for Materials Science, 1-2-1 Sengen, Tsukuba, Ibaraki 305-0047, Japan

**ABSTRACT**

The transverse thermoelectric generation and cooling performances in a thermopile module composed of recently developed $SmCo_5/Bi_{0.2}Sb_{1.8}Te_3$ artificially tilted multilayers are evaluated quantitatively. When a large temperature difference of 405 °C is applied to the $SmCo_5/Bi_{0.2}Sb_{1.8}Te_3$-based module, the open-circuit voltage and output power reach 0.51 V and 0.80 W, respectively. The corresponding maximum power density is 0.16 W/cm$^2$, even if the power is normalized by the device area including areas that do not contribute to the power generation, such as epoxy resin, electrodes, and insulating layers. The maximum energy conversion efficiency for our module in this condition is experimentally determined to be 0.92%. Under the cooling operation, the same module exhibits the maximum temperature difference of 9.0 °C and heat flow at the cold side of 1.6 W. Although these values are lower than the ideal thermoelectric performance expected from the material parameters due to the imperfections associated with modularization, the systematic investigations reported here clarify a potential of the $SmCo_5/Bi_{0.2}Sb_{1.8}Te_3$ artificially tilted multilayers as thermoelectric generators and cooling devices.

**Keywords**: transverse thermoelectric generation, electronic cooling, thermoelectric module, permanent magnet

## 1. Introduction

Thermoelectric conversion is a promising technology that enables direct conversion between heat and electricity, offering applications in waste heat recovery, power generation, and solid-state cooling [1,2]. Most of the current thermoelectric devices rely on the longitudinal thermoelectric effects, i.e., the Seebeck and Peltier effects, where heat and charge currents are interconverted in the same direction. In contrast, transverse thermoelectric conversion exploits alternative mechanisms in which heat and charge currents are interconverted in the orthogonal directions, offering unique advantages in device design and applications [3-6]. The transverse thermoelectric conversion has been observed in conductors under magnetic fields, magnetic materials with spontaneous magnetization, and conductors exhibiting anisotropic transport properties. The transverse thermoelectric conversion due to the anisotropic carrier conduction is driven by the off-diagonal Seebeck effect, which occurs in single crystals showing unipolar/ambipolar anisotropy [7-14] and in artificially tilted multilayers (ATMLs) comprising two different materials alternately and obliquely stacked [15-25].

Recently, as a promising transverse thermoelectric material, a multifunctional composite magnet (MCM) has been proposed [25]. Ando *et al*. have demonstrated that MCM comprising $SmCo_5/Bi_{0.2}Sb_{1.8}Te_3$ (BST) ATML exhibits excellent transverse thermoelectric power generation, free from performance degradation due to interfacial electrical and thermal resistances, and permanent magnet features simultaneously. The figure of merit for the off-diagonal Seebeck effect in $SmCo_5$/BST-based MCM was estimated to be $z_{xy}T$ = 0.20 at room temperature. Using the MCM-based thermopile module, the thermoelectric power generation of 204 mW at a temperature difference of 152 K has been demonstrated; the corresponding power density normalized by a heat transfer area and temperature gradient is record-high among transverse thermoelectric modules, while the energy conversion efficiency is not experimentally determined. Furthermore, the permanent magnet features of MCM enables easy implementation of the module on a heat source/bath made of magnetic materials through the magnetic attractive force and efficient thermal energy harvesting through reduced contact thermal resistances between the module and heat source/bath. Based on these results and the Onsager reciprocal relation, $SmCo_5$/BST-based MCM is expected to exhibit an excellent transverse thermoelectric cooling performance through the off-diagonal Peltier effect, but its cooling performance is yet to be evaluated. The reduced contact thermal resistance through the magnetic attractive force from MCM should be useful in the cooling operation in the same manner as the power generation operation.

In this study, we have comprehensively investigated the transverse thermoelectric conversion performances for the MCM-based thermopile module consisting of $SmCo_5$/BST ATMLs. First, using an MCM-based module with a larger size than that used in Ref. [25], we experimentally estimated the open-circuit voltage, output power, and energy conversion efficiency under large temperature differences. We also measured the maximum temperature difference and coefficient of performance (COP) for cooling operation using the same module and discussed the relationship between the maximum temperature difference and $z_{xy}T$ in transverse thermoelectric cooling based on the phenomenological formulation. This work further confirms the significant potential of MCM for thermal energy harvesting and thermal management applications.

## 2. Experimental details

### *2.1. Sample preparation*

The MCM-based module was constructed in the similar manner to the procedures in Ref. [25]. The alternately stacked $SmCo_5$/BST multilayer slab was prepared by a spark plasma sintering method under a pressure of 30 MPa at 450 °C for 30 min, where the thickness of each layer is 0.5 mm. The microstructure analysis for the $SmCo_5$/BST junction was also reported in Ref. [25]. The $SmCo_5$/BST ATMLs, MCM elements, with a length of 15.0 mm, width of 1.5-1.8 mm, thickness of 7.2-7.3 mm, and tilt angle $\theta$ of 25° was cut from the multilayer slab, where Cerasolzer #297 (Kuroda Techno Co., Ltd.) electrodes with a melting point of 297 °C were attached to the 1.5-1.8 × 7.2-7.3 mm surfaces of the elements using an ultrasonic soldering method. The present MCM-based module consists of the 16 MCM elements alternately stacked in the width direction with opposite $\theta$ and intermediated by thin insulating papers and glue with a heat resistance of 1100 °C (HJ-112, Cemedine Co., Ltd.), while the module used in Ref. [25] has 14 elements. The MCM elements were electrically connected in series to form a zigzag thermopile circuit and enameled Cu wires were attached to the ends of the circuit by melting Cerasolzer #297. The total length, width, and thickness of the MCM-based module except for the Cu wires are around 17.5, 27.8, and 7.4 mm, respectively, where the module size is slightly larger than the ideal one expected from the element size due to small stacking misalignment. Here, the temperature gradient is applied along the thickness direction. To increase the mechanical strength of the module, the area around the electrodes and the base of the Cu wires were hardened with epoxy resin having a heat resistance of 340 °C (Duralco 4703, Cotronics Corp.). Finally, to flatten the heat transfer surfaces, the 17.5 × 27.8 mm surfaces of the module were also covered with a thin layer of the same resin. As described above, the components with the lowest heat resistance in our MCM-based module are the electrodes. However, during the measurements of thermoelectric power generation, only the top surface is exposed to the highest temperature, and the MCM-based module can operate stably even at a hot side temperature of $T_H$ ~ 400 °C, as long as a cold side temperature is kept at $T_C < 0$ °C. In fact, before performing the power generation measurements shown below, the same module was subjected to two heating/cooling cycles with $T_H$ ranging from room temperature to 400 °C, and no signs of degradation, e.g., the change in the internal resistance of the module, were observed throughout these repeated cycles. In longitudinal thermoelectric modules, junctions between the elements and electrodes are exposed to high temperatures, which causes degradation, whereas MCM-based modules have the advantage of high durability as described above.

### *2.2. Evaluation of power generation characteristics*

The thermoelectric power generation properties, such as the output power and conversion efficiency, of the constructed MCM-based module were evaluated using an in-house developed thermoelectric module characterization system [26]. The configuration for evaluating the power generation characteristics of the module is shown in Figure 1(a). A temperature difference $\Delta T$ was applied to the module by inputting heat $Q_H$ to the hot side of the module using a heater and dissipating heat $Q_C$ from the cold side of the module to a cooling stage whose temperature was controlled by Peltier elements and a chiller. $T_H$ and $T_C$ were measured by thermocouples inserted into holes in heat baths of AlN ceramic substrates with an area of 28 × 28 mm$^2$ and thickness of 2 mm attached via metal-oxide-filled silicone compound (Thermal joint type C, Kataoka Senzai) to the hot and cold side surfaces of the module. In the previous study, the actual temperature gradient inside the module was measured using an infrared camera [25], whereas in this study, $T_H$ and $T_C$ were measured in two heat baths with proper thermal contact to outer surfaces of the module, which is commonly adopted in thermoelectric module evaluations [27,28]. The evaluation was performed in a vacuum of < $10^{-1}$ Pa. The module and AlN ceramic plates were pressurized at 0.4 MPa to ensure proper thermal contact between these components with suitable thermal interfaces of a graphite sheet with a thickness of 0.127 mm (Grafoil, NeoGraf Solutions) and silicone grease (KS-613, Shin-Etsu Silicone). By controlling $Q_H$ and $Q_C$, $\Delta T$ (= $T_H - T_C$) of 115, 215, 310, 405 °C were applied with $T_C$ ranging from −15 °C to −5 °C. The module output power $P_O$ at a certain $\Delta T$ was calculated by the product of the load current $I_L$ regulated by an electronic load (PLZ164WA, Kikusui Electronics) and the output voltage $V_O$ measured by a digital multimeter (Keithley 6500, Tektronix). The thermoelectric conversion efficiency $\eta$ is defined as $P_O/Q_H$. Here, $Q_H$ was estimated by the sum of $Q_C$ and $P_O$ taking into account the energy balance, since the accurate measurement of $Q_H$ is difficult due to heat leakage caused by radiation at high temperatures. The $Q_C$ values were determined by a heat flow meter located on an oxygen-free Cu block with known thermal conductivity attached to the cold side of the module. As shown in Figure 1(a), six tiny Pt resistance thermometers were embedded in the Cu block, and $Q_C$ was determined from the measured temperature drop along the heat flow direction and the dimensions of the Cu block, considering the one-dimensional Fourier's law. The AC resistance of the module $R_{AC}$ was measured by an inductance-capacitance-resistance meter (BT3562, HIOKI) at each $\Delta T$.

### *2.3. Evaluation of electronic cooling characteristics*

The thermoelectric cooling properties, such as the maximum temperature difference and COP, of the MCM-based module were evaluated using the same equipment as for power generation. The configuration for evaluating the cooling characteristics of the module is shown in Figure 1(b). While most of the components were common between the power generation and cooling evaluations, the major difference was that the heat flow meter was inserted between the heater and module. In addition, $T_H$ and $T_C$ are defined on the cooling stage and heater sides of the module, respectively, which is the opposite of the case for power generation. The heat baths were fixed to the module surfaces with epoxy resin to stabilize the thermal contact, since the risk of module breakdown due to the application of high temperatures is low in the evaluation of the cooling characteristics. Figure 2 shows photographs of the setup for evaluating the cooling characteristics. COP is defined as the ratio of the amount of heat removed by the cold side of the module $Q_C$ to the power input to the module $P_I$, and is expressed as COP = $Q_C/P_I$. Here, $Q_C$ is determined by the heat flow meter and $P_I$ is calculated by the product of the input current $I_I$ and voltage $V_I$ applied to the module. At each input current $I_I$ = 1, 4, 5, 6, 7, 8, and 10 A, $T_H$ was maintained at 50 °C and $V_I$ was measured at several $T_C$ controlled by the heater. The degree of vacuum, the pressure applied to the components, and the thermal interfaces were the same as those during the power generation evaluation.

## 3. Results and discussion

### *3.1. Power generation*

Figure 3 shows the evaluation results of the thermoelectric power generation characteristics for the MCM-based module. Figure 3(a) shows the $I_L$ dependence of $V_O$ and $P_O$ on the left and right axes, respectively. The open-circuit voltage $V_{OC}$ at $\Delta T$ = 115 °C observed here is comparable to but slightly smaller than that at a temperature difference of 106 °C reported in Ref. [25]. This difference can be caused by various factors, such as the small difference in the position where the temperature difference is measured, electrode quality, and average temperature of the module. The $V_{OC}$ value and maximum output power $P_{O,max}$ increased with $\Delta T$ and showed maximum values of 0.51 V and 0.80 W at $\Delta T$ = 405 °C, respectively. This maximum power was about four times higher than 204 mW in the previous report [25]. In this condition, the maximum power density and that normalized by the temperature gradient were 0.16 W/cm$^2$ and 0.055 mW/cm$^2$·(mm/K)$^2$, respectively, which were comparable to those of commercial longitudinal thermoelectric modules [29-31]. Here, we note that the device area used for the normalization includes the areas of the epoxy resin, electrodes, and insulating layers that do not contribute to the power generation; if the power is normalized by the total area of the ATML elements, the maximum power density becomes approximately 20% larger. Figure 3(b) shows $Q_C$ and $\eta$ on the left and right axes, respectively. $\eta$ also increased with $\Delta T$, and the maximum efficiency $\eta_{max}$ reached 0.92% when $\Delta T$ and $I_L$ were 405 °C and 3.1 A, respectively. Although this $\eta_{max}$ value is still smaller than that for longitudinal modules [32-41], it is a promising value among transverse modules. The device figure of merit was estimated to be 0.063 for the MCM-based module using the measured $\eta_{max}$ and the relationship between $\eta_{max}$ and $z_{xy}T$ in transverse thermoelectric conversion [42]. The discrepancy between this device figure of merit and the material $z_{xy}T$ reported in Ref. [25] is discussed in the next subsection. Figure 3(c) shows the $\Delta T$ dependence of $P_{O,max}$ and $\eta_{max}$ on the left and right axes, respectively. Figure 3(d) shows the $\Delta T$ dependence of $T_C$ and $R_{AC}$ on the left and right axes, respectively. At $\Delta T$ = 115 °C and 215 °C, $T_C$ could be maintained at −15 °C; however, due to insufficient cooling capacity, $T_C$ increased with $\Delta T$ and reached $T_C$ = −5 °C at $\Delta T$ = 405 °C.

### *3.2. Electronic cooling*

Figure 4 shows the evaluation results of the cooling characteristics for the same MCM-based module. Figure 4(a) and 4(b) show the $\Delta T$ dependence of $Q_C$ and $P_I$ at each $I_I$, respectively. By linear fitting the $Q_C$ data in Figure 4(a), the $I_I$ dependence of the maximum temperature difference $\Delta T_{max}$ ($\Delta T$ at $Q_C$ = 0 W) and the maximum heat flow at the cold side $Q_{C,max}$ ($Q_C$ at $\Delta T$ = 0 °C) were determined as shown on the left and right axes of Figure 4(c), respectively. The values of $\Delta T_{max}$ and $Q_{C,max}$ were estimated to be 9.0 °C and 1.6 W at $I_I$ = 6.8 A and 7.1 A, respectively. Kyarad *et al.* demonstrated $\Delta T_{max}$ > 20 °C at an extremely large current of 40 A in a single element of Pb/$Bi_2Te_3$ ATML [43]. The applied current can be reduced by constructing the thermopile module with multiple elements as described in this study. However, in exchange, the performance can be degraded due to experimental imperfections including small but finite contact resistances at the electrodes and misalignment of the elements (note that such imperfections are unavoidable in laboratory-level verification, but can be minimized once mass-production processes are established). In addition, $\Delta T_{max}$ can be increased by introducing an infinite-stage structure as often demonstrated for Ettingshausen coolers [44-46], while it is difficult to introduce such structures in the thermopile module. Figure 4(d) shows the $I_I$ dependence of COP determined from $Q_C$ and $P_I$ at the certain $\Delta T$ by linear fitting the $P_I$ data in Figure 4(b).

Here, we analyze $\Delta T_{max}$ for the present MCM-based module. Following the method of Ref. [47], the maximum attainable temperature difference is calculated as

$$\Delta T_{\mathrm{max}} = \frac{1}{2} z_{xy} {T_{\mathrm{C}}}^2, \qquad (1)$$

where $z_{xy} = {S_{xy}}^2/\rho_{xx}\kappa_{yy}$, $S_{xy}$ is the off-diagonal Seebeck coefficient, $\rho_{xx}$ is the electrical resistivity along the applied charge current direction (*x* direction), $\kappa_{yy}$ is the thermal conductivity along the generated temperature gradient (*y* direction), and we assume the isothermal boundary condition along the *x* direction as the device geometry adopted in this study is more like Figure 1(d) of Ref. [11] than Figure 1(c). This equation shows that the simple relationship between the maximum temperature difference and figure of merit is applicable not only to the longitudinal thermoelectric conversion but also to the transverse thermoelectric conversion. On the other hand, for the transverse thermoelectric conversion operating with the Ettingshausen effect in which time-reversal symmetry is broken, the relationship between $\Delta T_{max}$ and the figure of merit is Equation (1) with $T_C$ replaced by $T_H$ [47,48]. When the experimentally obtained $\Delta T_{max}$ value is substituted into Equation (1), $z_{xy}T_C$ is estimated to be 0.057.

The device figure of merit calculated from $\eta_{max}$ of the power generation and the $z_{xy}T_C$ value calculated from $\Delta T_{max}$ under the cooling operation were close, although the definitions of temperature were different. These values are significantly reduced from $z_{xy}T = 0.20$ for $SmCo_5$/BST ATML reported in Ref. [25] due to the following three reasons. Firstly, the device figure of merit determined from the module performance generally decreases from $z_{xy}T$ estimated from ${S_{xy}}^2/\rho_{xx}\kappa_{yy}$ due to the electrical and thermal shunting effects at the boundaries to the electrodes and heat source/bath [16,18] as well as the bonding condition between the ATML element and electrodes. Thus, there remains room to further improve the power generation and cooling performances by optimizing the electrical and thermal boundaries for MCM-based modules. Secondly, when a large temperature difference is applied, a part of the material in the module is at a temperature outside the optimal operating temperature, and the device figure of merit effectively decreases. Thirdly, the power generation and cooling performances could be underestimated because $T_H$ and $T_C$ were measured outside of the module in this study, while $\nabla T$ inside the module was measured in the previous study.

### *3.3. Numerical simulation*

Finite element method simulations were conducted using COMSOL Multiphysics 6.3 to reproduce the measured results of power generation characteristics. The temperature dependence of the Seebeck coefficient, electrical resistivity, and thermal conductivity of BST and $SmCo_5$ reported in Ref. [25] were used for the simulations. To investigate the factors contributing to the reduced power generation performance of the MCM-based module, the simulations at $\Delta T = 405$ K were systematically carried out in a configuration close to the experimental setup with adjusting three parameters. First, the electrode thickness of the ATML elements was adjusted to align the simulated heat flow with the experimental results. Second, the thermal conductivity of the insulating layer between the AlN ceramic plates and ATML elements was modified to account for potential thermal contact resistance. Third, the influence of the electrical contact resistance was taken into consideration by modifying the conductivity of the lead wires connecting the electrodes of each ATML element. Figures 5(a) and 5(b) present the 3D simulation results of the temperature and electric potential distributions, respectively. Figure 5(c) shows the experimental and simulated results of $V_O$ and $P_O$ as functions of $I_L$ at $\Delta T = 405$ °C. The simulated temperature difference across the ATML elements at $I_L = 0$ A was 323 °C, corresponding to approximately 80% of the temperature difference between the AlN ceramic plates due to large thermal contact resistance of the insulating layer. In contrast, the slope of the *I*-*V* characteristics was consistent with the experimental result when an additional contact resistance of 26 mΩ was assumed for the lead wires. The experimentally obtained $R_{AC}$ at $\Delta T = 405$ °C was 78 mΩ, suggesting that the contact resistance could constitute approximately 33% of the total resistance. Figure 5(d) shows the experimental and simulated results of $Q_C$ and $\eta$ as functions of $I_L$. The simulated heat flow bypassing the electrodes of the ATML elements at $I_L$ = 0 A was 18 W, corresponding to approximately 21% of the experimentally measured $Q_C$ of 84 W. These factors may have caused the $z_{xy}T$ value estimated from the power generation efficiency to be significantly lower than that obtained from the evaluation of an individual element. On the other hand, our efficiency evaluation is based on heat flow measurements on the cold side and therefore does not take into account radiation losses from the side surfaces of the module. Assuming an emissivity of 1 on the module side surfaces as a worst-case condition, the simulated radiative loss reached up to 3.2 W, possibly leading to an overestimation of the conversion efficiency by up to 4%. Although the MCM-based module was evaluated under a vacuum of $<10^{-1}$ Pa, our previous findings indicate that convective heat loss at this pressure level is negligible compared to the total heat flow through the module. Achieving proper thermal contact was difficult in transverse modules because they do not allow the use of conductive materials such as graphite sheets commonly used to attach longitudinal modules to heat baths. For longitudinal modules, suitable thermal contact can be achieved by eliminating the hot-side ceramic substrate and directly attaching a graphite sheet to the hot-side electrode [26]. In contrast, for transverse modules based on ATML, where the thermoelectric voltage is generated in the in-plane direction, conductive graphite sheets cannot be directly used at the hot-side contact because they may shunt each ATML element electrically. Instead, an insulating material, such as epoxy resin, must be used, which may potentially degrade the thermal contact. Therefore, further improvements can be expected by obtaining better thermal contact with heat baths, such as using insulating coatings with high thermal conductivity. Nevertheless, the transverse thermoelectric conversion performance estimated here is far superior to that obtained by the anomalous Nernst and Ettingshausen effects in magnetic materials, confirming the usefulness of $SmCo_5$/BST-based MCM.

### *3.4. Trial for construction of larger-area modules*

The construction of larger-area modules is important to obtain larger maximum output power and cooling performance. To demonstrate this, we fabricated a larger MCM-based module through the same procedures and performed the same evaluation of the power generation performance. The new module is composed of 40 $SmCo_5$/BST ATML elements and has an area of ~33 × 31 mm. Therefore, under the same temperature gradient, we expected an output power about twice as great as that of the smaller module used in the above experiments. However, the maximum output power normalized by the square of the temperature gradient for the larger module was only ~1.2 times larger than that for the smaller one. This suggests that the performance degradation due to modularization imperfections becomes more pronounced in larger modules with the larger number of junctions, highlighting the importance of junction-less modules.

## 4. Conclusions

In this study, we evaluated the transverse thermoelectric power generation and cooling performances of a MCM-based thermopile module composed of $SmCo_5$/BST ATML having the experimentally determined figure of merit of 0.20 at room temperature. Although the performances were observed to be lower than expected from the material properties due to the modularization, our module showed excellent features as a transverse thermoelectric convertor in both power generation and cooling operations. To construct a large-area module with a higher output that fully utilizes the potential of MCM, it is essential to optimize electrodes and improve cutting and stacking accuracies of MCM elements. If these modularization techniques are established, MCM will bring innovation to the development of next-generation thermal management technologies.

## Acknowledgements

The authors thank K. Suzuki, M. Isomura, H. Ohshima, and K. Aoyama for technical support.

## Disclosure statement

The authors declare that they have no known competing financial interests or personal relationships that could have appeared to influence the work reported in this paper.

## Funding

This work was supported by ERATO "Magnetic Thermal Management Materials" (No. JPMJER2201) from Japan Science and Technology (JST), Grants-in-Aid for Scientific Research (KAKENHI) (No. 24K17610) from Japan Society for the Promotion of Science (JSPS), and NEC Corporation.

## Data availability statement

The data that support the findings of this study are available from the corresponding authors upon reasonable request.

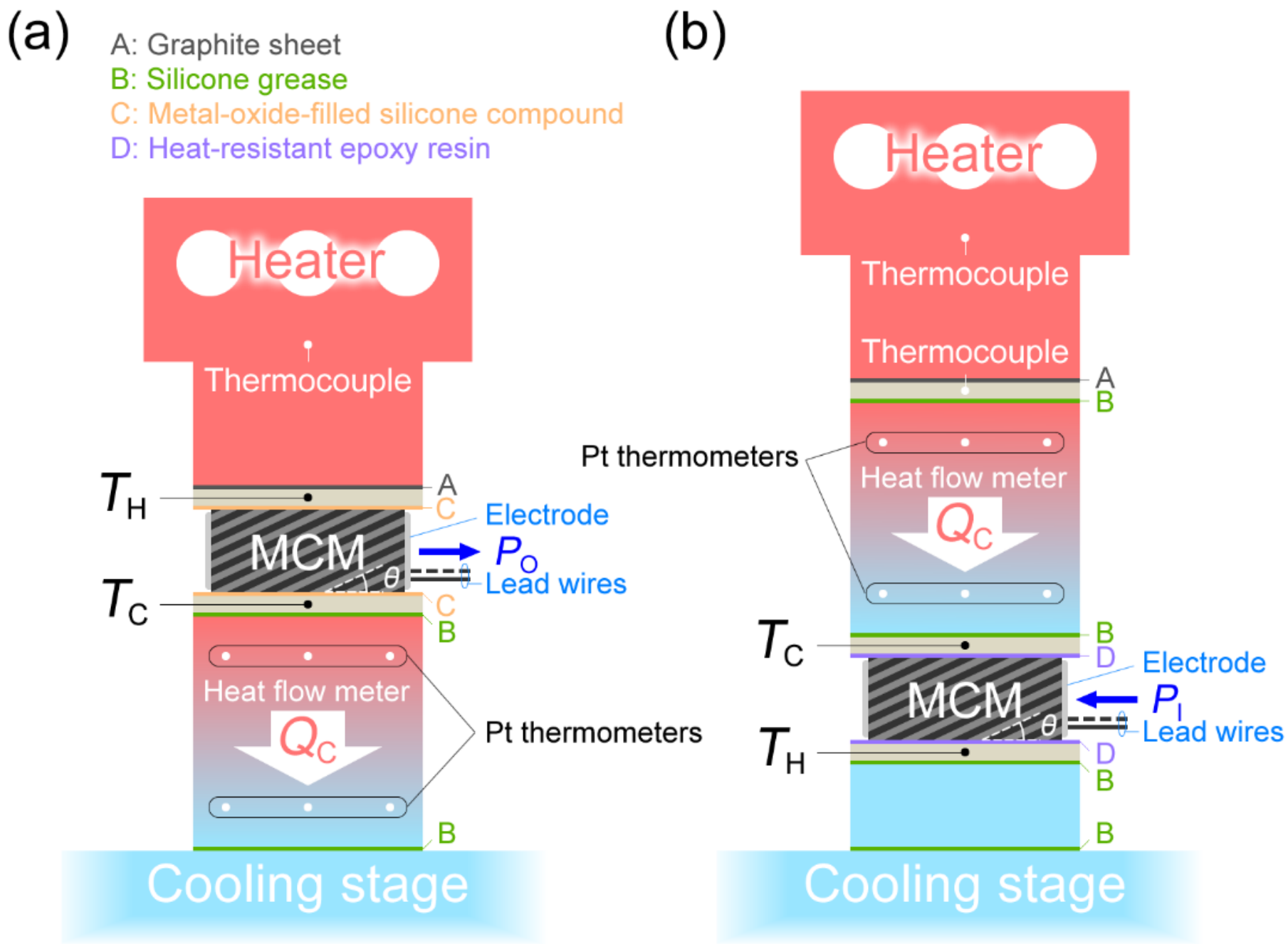

**Figure 1. Schematics of evaluation system for MCM-based module.** (a) Configuration for evaluating power generation characteristics. (b) Configuration for evaluating cooling characteristics.

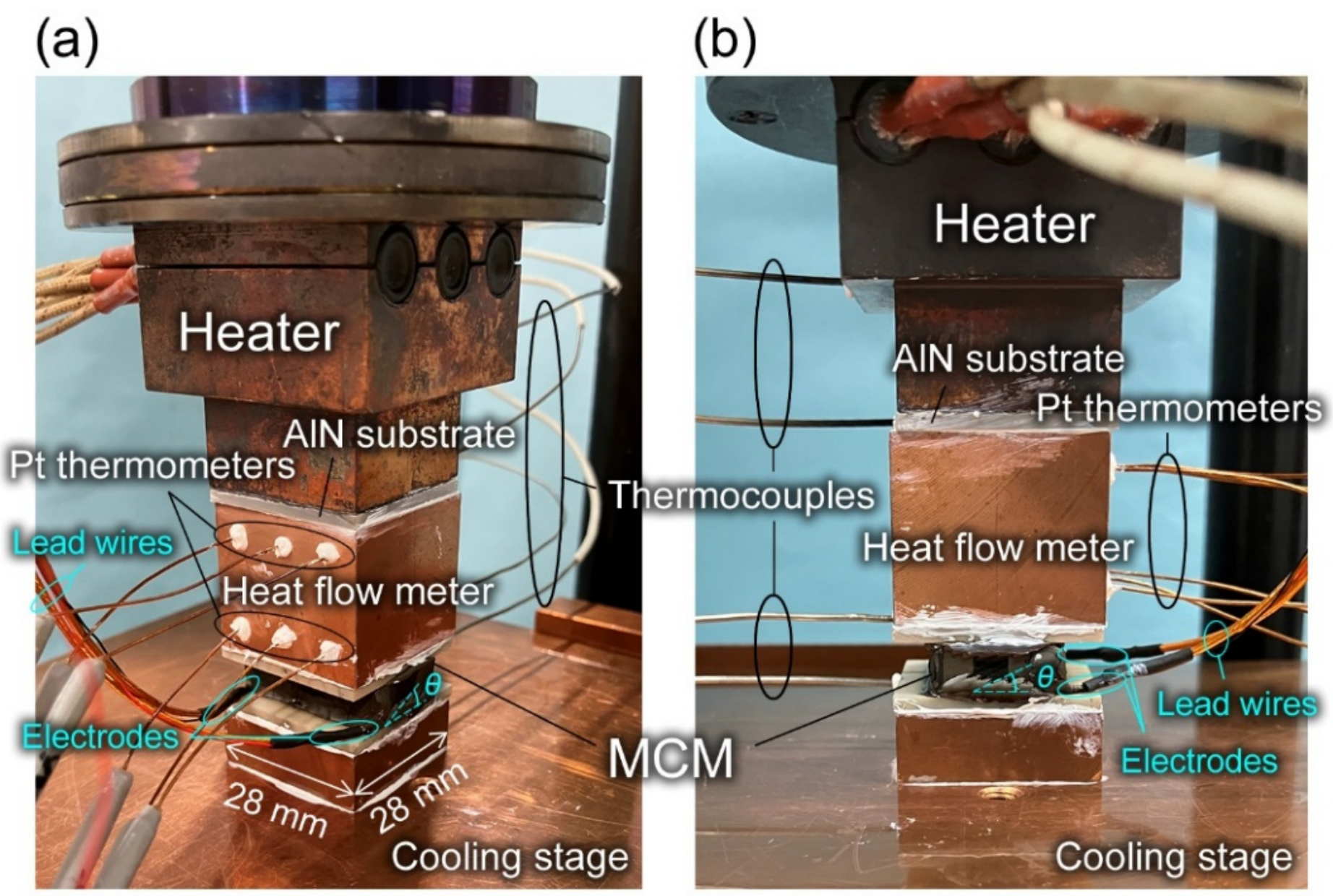

**Figure 2. Photographs of the setup for evaluating cooling characteristics of MCM-based module.** (a) Bird's-eye view. (b) Side view.

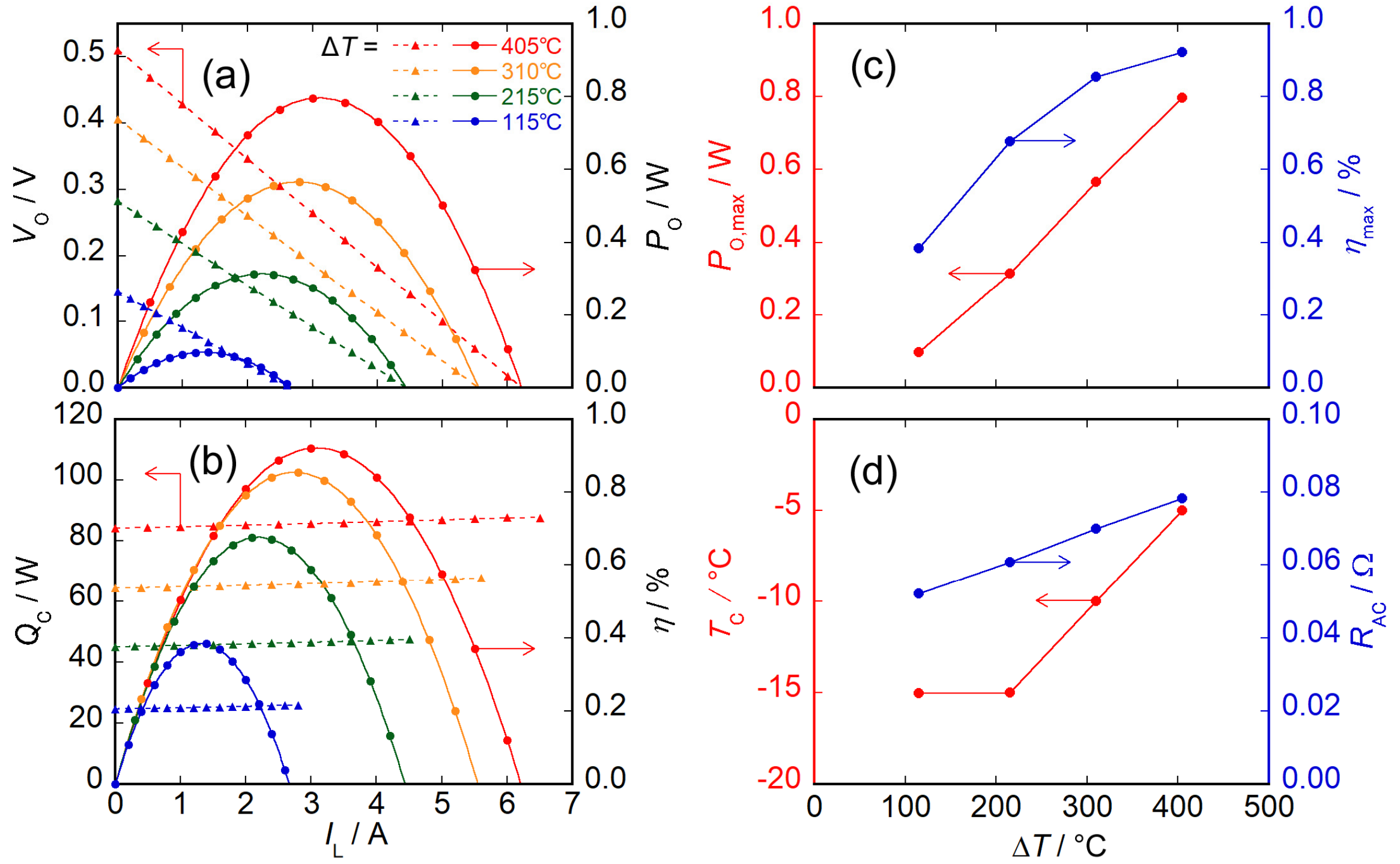

**Figure 3. Evaluation results of thermoelectric power generation characteristics for MCM-based module.** (a) $I_L$ dependence of $V_O$ and $P_O$. (b) $I_L$ dependence of $Q_C$ and $\eta$. (c) $\Delta T$ dependence of $P_{O,max}$ and $\eta_{max}$. (d) $\Delta T$ dependence of $T_C$ and $R_{AC}$.

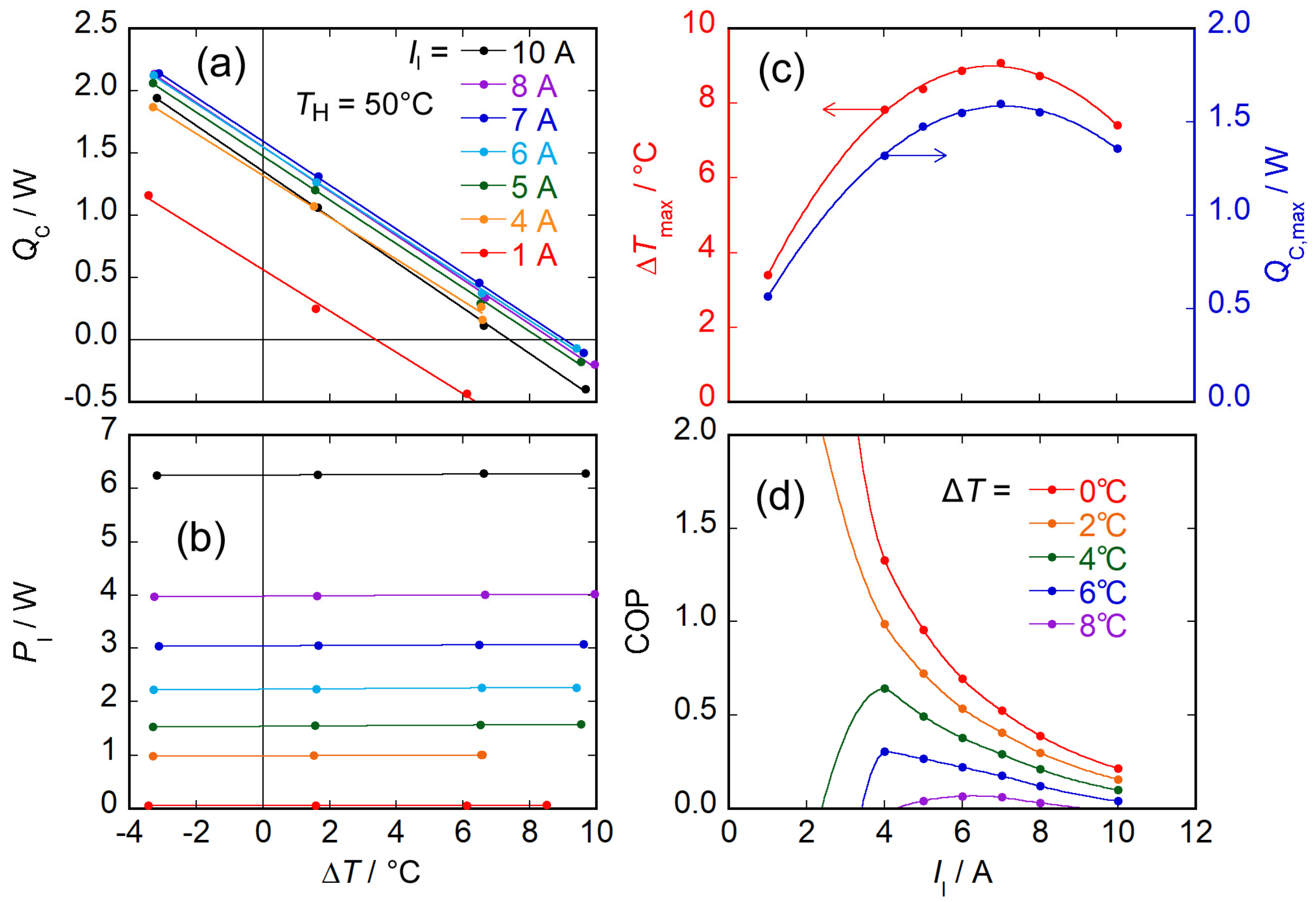

**Figure 4. Evaluation results of cooling characteristics for MCM-based module.** (a) $\Delta T$ dependence of $Q_C$ in each $I_I$. (b) $\Delta T$ dependence of $P_I$ in each $I_I$. (c) $I_I$ dependence of $\Delta T_{max}$ and $Q_{C,max}$. (d) $I_I$ dependence of COP at each $\Delta T$.

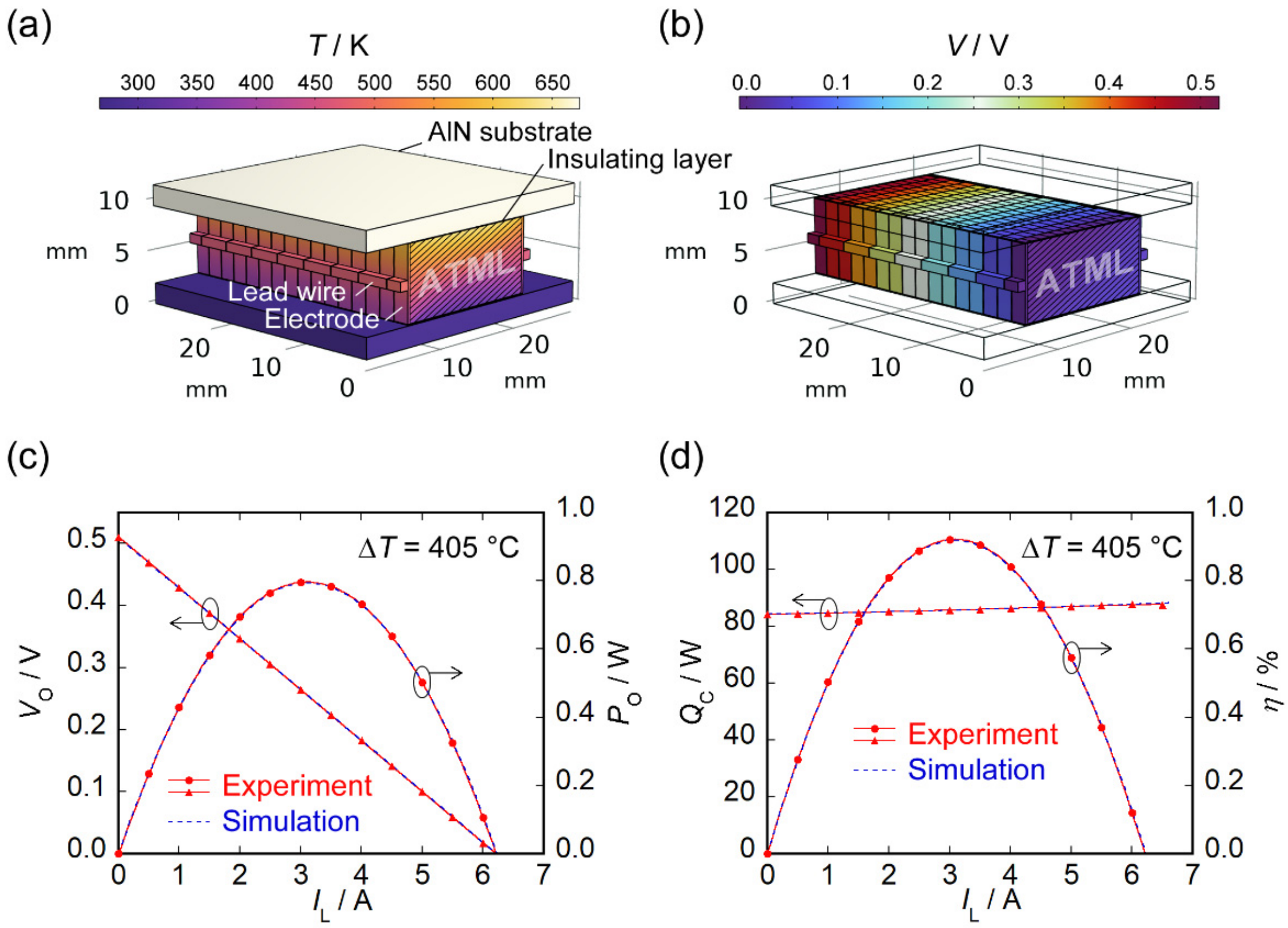

**Figure 5. Finite element method simulations of thermoelectric power generation characteristics of MCM-based module.** (a) 3D simulation results of the temperature distribution. (b) 3D simulation results of the electric potential distribution. (c) Experimental and simulated results of the $I_L$ dependence of $V_O$ and $P_O$ at $\Delta T = 405$ °C. (d) Experimental and simulated results of the $I_L$ dependence of $Q_C$ and $\eta$ at $\Delta T = 405$ °C.